\documentclass[aps,prb,reprint,superscriptaddress,nofootinbib,showkeys]{revtex4-2}

\usepackage{amsmath,amssymb,bm}
\usepackage{graphicx}
\usepackage{booktabs,multirow,array}
\usepackage{xcolor}
\usepackage[colorlinks=true,linkcolor=blue,citecolor=blue,urlcolor=blue]{hyperref}
\usepackage{newunicodechar}

\newcommand{\tcellc}[2]{\parbox[t]{#1}{\centering #2}}
\newunicodechar{Γ}{\ensuremath{\Gamma}}
\newunicodechar{α}{\ensuremath{\alpha}}
\newunicodechar{π}{\ensuremath{\pi}}
\newunicodechar{σ}{\ensuremath{\sigma}}
\newunicodechar{∆}{\ensuremath{\Delta}}
\newunicodechar{∪}{\ensuremath{\cup}}
\newunicodechar{⊥}{\ensuremath{\perp}}
\newunicodechar{×}{\ensuremath{\times}}
\newunicodechar{±}{\ensuremath{\pm}}
\newunicodechar{·}{\ensuremath{\cdot}}
\newunicodechar{⋅}{\ensuremath{\cdot}}
\newunicodechar{°}{\ensuremath{^\circ}}
\newunicodechar{Å}{\AA{}}

\begin{document}

\renewcommand{\figurename}{Fig.}
\renewcommand{\tablename}{Table}

\title{Symmetry-constrained coexistence of charge-three and charge-one Weyl points in nonmagnetic crystals}

\author{Ze-Xin Xue}
\author{Ke-Xin Pang}
\author{Yun-Yun Bai}
\author{Yanfeng Ge}
\author{Yan Gao}
\email{yangao9419@ysu.edu.cn}
\affiliation{State Key Laboratory of Metastable Materials Science and Technology and Hebei Key Laboratory of Microstructural Material Physics, School of Science, Yanshan University, Qinhuangdao 066004, China}

\date{\today}

\begin{abstract}
High-charge Weyl semimetals (WSMs), particularly those hosting charge-three Weyl point (C-3 WPs), have attracted considerable interest owing to their unconventional topological properties and enhanced boundary phenomena. However, ideal C-3 WSMs remain rare, and the symmetry-constrained coexistence of charge-three and charge-one Weyl point (C-1 WPs) in nonmagnetic crystals remains largely unexplored. Here, we investigate a representative eight-node Weyl configuration, consisting of two C-3 WPs and six C-1 WPs with opposite chirality, corresponding to two identical $\{3,-1,-1,-1\}$ irreducible Weyl molecules. By combining topological charge conservation with space-group symmetry analysis, we systematically classify all 230 space groups and identify 12 and 23 candidate groups without and with spin--orbit coupling (SOC), respectively. First-principles calculations without SOC reveal that the boron allotropes TBR-B$_{48}$ and chiral OBD-B$_{48}$ realize this configuration, hosting hourglass-type and non-hourglass-type C-3 WPs, respectively. Surface-state calculations reveal six Fermi-arc branches at the shared projection of the charge-three pair and three at each individual projection. Reversing the structural chirality of OBD-B$_{48}$ reverses the signs of all Weyl charges, establishing a direct correspondence between crystal and momentum-space chirality. Furthermore, breaking the protecting $C_6$ symmetry in TBR-B$_{48}$ splits each C-3 WP into three C-1 WPs and reconstructs the surface Fermi arcs. Our work establishes symmetry-based guidelines for designing and searching for nonmagnetic WSMs hosting charge-three and charge-one Weyl-point coexistence.
\end{abstract}

\maketitle

\section{Introduction}\label{sec:introduction}
Weyl semimetals (WSMs) host isolated band crossings that act as monopoles or antimonopoles of Berry curvature in momentum space, with the quantized topological charge characterized by the Chern number $C$ defined on a closed surface enclosing each Weyl point (WP)~\cite{Armitage2018,Wan2011,Burkov2018,Yan2017}. While conventional WPs with $|C|=1$ have been extensively studied~\cite{Weng2015,TaAs,TaAsSurface,TypeII,MagneticWeyl,MXBT,HgTeIdeal}, crystalline symmetries can stabilize high-Chern-number WPs carrying multiple units of topological charge, including C-2, C-3, and C-4 WPs~\cite{multiweyl,C4Weyl,QuadrupleWeyl,ThreeNodeWeylComplex,SinglePairChargeTwoBoron,DoubleWeylCarbon,Tsirkin,WoodpileWeyl,TripleHourglass,HourglassPhonon,Li2CO4,MaxChargeWeyl}. In particular, rotation symmetries provide a fundamental mechanism for realizing high-charge WPs by constraining the lowest-order interband coupling near band crossings~\cite{multiweyl,Tsirkin}. The discovery of these unconventional WPs has expanded the landscape of topological quasiparticles beyond elementary charge-one nodes and revealed new connections between crystalline symmetry and momentum-space topology.

Among high-Chern-number Weyl systems, C-3 WPs are of particular interest because their multiple units of Berry flux generate distinctive topological phenomena, including multiple long Fermi arcs~\cite{multiweyl,Bradlyn2016Beyond}, enhanced quantized circular photogalvanic effects~\cite{CPGETheory,CPGE,MultifoldOptics}, and unconventional interaction-driven quantum critical behaviors~\cite{TripleWeylQCP,CorrelatedTripleWeyl}. However, despite these intriguing properties, realizing clean C-3 Weyl fermions in electronic systems remains challenging. In contrast to the relatively abundant realizations of hourglass C-3 Weyl phonons in bosonic systems~\cite{HourglassPhonon,TripleHourglass,WeylDipolePhonon,ChargeThreeAcoustic,UnpairedWeylPhonon}, fermionic C-3 WPs are strongly restricted by band filling, Fermi-level alignment, and spin--orbit coupling (SOC). Existing electronic realizations mainly rely on two strategies. One employs fully spin-polarized magnetic systems, such as BaNiIO$_6$ and SrNiIO$_6$~\cite{BaNiIO6,SrNiIO6}, but requires specific magnetic structures to enable isolated high-charge crossings~\cite{La3AlCrS7}, while spin-orbit-coupling (SOC) effects may modify the symmetry protection of these band crossings. The other utilizes light-element systems with negligible SOC, exemplified by Li$_2$CO$_4$~\cite{Li2CO4}, where the topology is determined predominantly by crystalline symmetry. Therefore, identifying clean nonmagnetic platforms for C-3 Weyl fermions remains an important challenge.

\begin{figure*}[t!]
	\centering
	\includegraphics[width=0.86\textwidth]{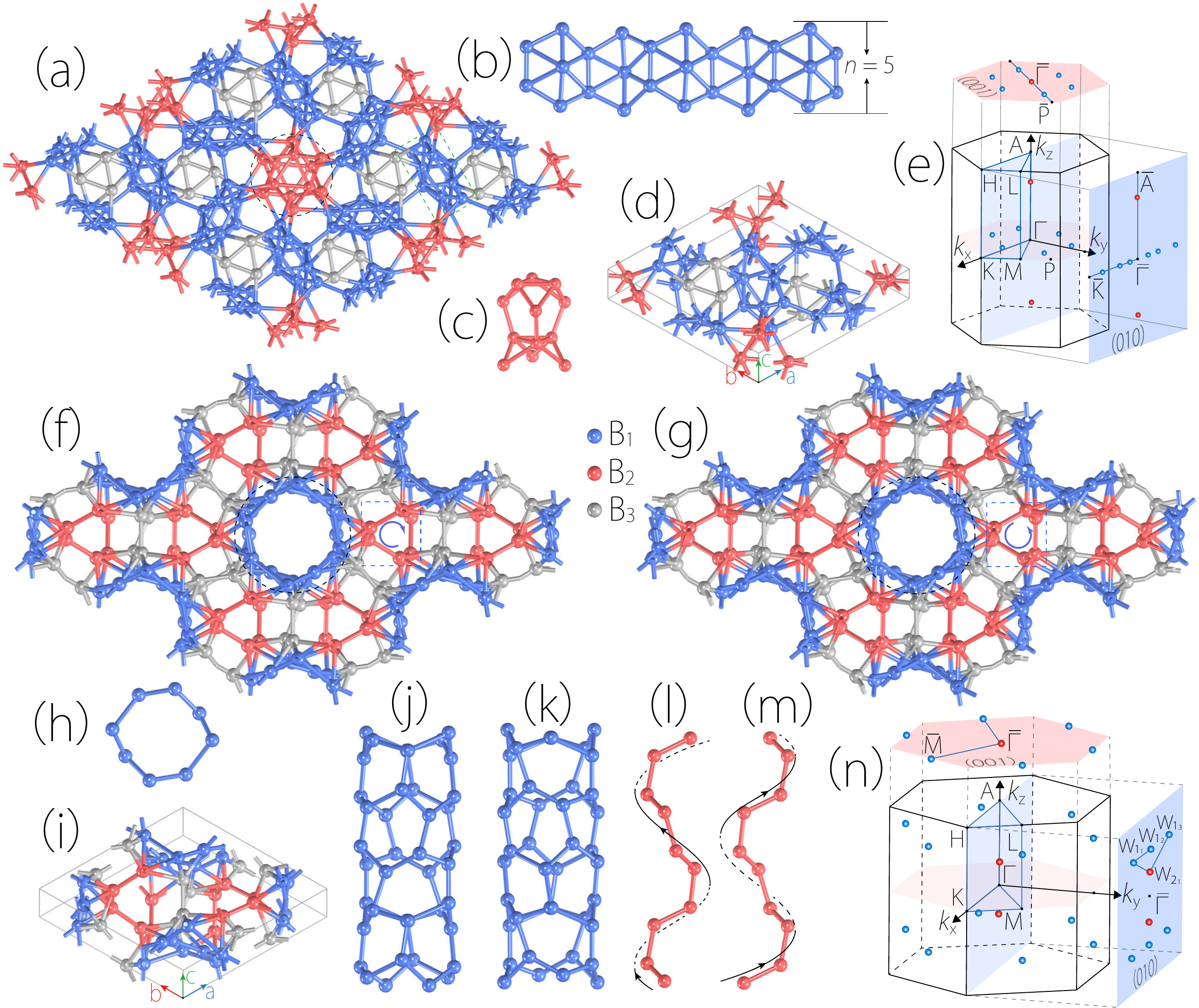}
	\caption{Crystal structures and Brillouin-zone (BZ) schematics of TBR-B$_{48}$ and the left- and right-handed OBD-B$_{48}$ structures. (a) Top view of the $2\times2\times1$ supercell of TBR-B$_{48}$. (b) Buckled boron nanoribbon composed of B$_1$ atoms and five atoms in width. (c) Side view of the structural unit marked by the black dashed region in (a). (d) Perspective view of the optimized primitive cell of TBR-B$_{48}$. (e) Three-dimensional bulk BZ of TBR-B$_{48}$ and its projections onto the (001) and (010) surfaces. (f) Top view of the $2\times2\times1$ supercell of \emph{l}-OBD-B$_{48}$. (g) Top view of the $2\times2\times1$ supercell of \emph{r}-OBD-B$_{48}$. (h) Top view of the buckled eight-membered ring forming the blue nanotube in OBD-B$_{48}$. (i) Perspective view of the optimized primitive cell of OBD-B$_{48}$. (j), (k) Side views of the buckled eight-membered ring nanotubes in \emph{l}- and \emph{r}-OBD-B$_{48}$, respectively. (l), (m) Triangular single-atom-wide dimer boron chains formed by helical rotation along the $c$ axis in \emph{l}- and \emph{r}-OBD-B$_{48}$, respectively. (n) Three-dimensional bulk Brillouin zone (BZ) of OBD-B$_{48}$ and its projections onto the two-dimensional BZs of the (001) and (010) surfaces, with high-symmetry points and lines labeled.}
	\label{fig:structure}
\end{figure*}

Beyond the realization of an isolated C-3 WP, the existence of high-charge Weyl fermions is fundamentally constrained by the global topology of the entire Brillouin zone (BZ). The Nielsen--Ninomiya no-go theorem requires the total chirality of all WPs in a lattice to vanish~\cite{NielsenNinomiya1,NielsenNinomiya2}. In a spinless time-reversal-symmetric system with $C_6$ rotational symmetry, C-3 WPs can appear along the $\Gamma$-$A$ line~\cite{multiweyl,Encyclopedia}, and the two time-reversal-related WPs located at $\mathbf{k}$ and $-\mathbf{k}$ possess the same chirality. Consequently, a pair of C-3 WPs with total charge $\pm6$ must be compensated by at least six C-1 WPs carrying the opposite total charge. This requirement naturally leads to a minimal charge-neutral configuration consisting of two identical $\{3,-1,-1,-1\}$ irreducible Weyl molecules (IWMs)~\cite{IWM}, each composed of one C-3 WP and three compensating C-1 WPs. Although several nonmagnetic C-3 Weyl materials have been reported~\cite{HourglassPhonon,TripleHourglass,Li2CO4,WeylDipolePhonon,ChargeThreeAcoustic,UnpairedWeylPhonon,BaNiIO6,SrNiIO6,La3AlCrS7}, ideal C-3 WSMs hosting the double $\{3,-1,-1,-1\}$ configuration, along with a systematic understanding of their crystallographic origin, symmetry constraints, and realizable space groups (SGs) remain lacking.

In this work, we combine space-group symmetry analysis, low-energy $k\cdot p$ modeling, and first-principles calculations to systematically investigate the double $\{3,-1,-1,-1\}$-IWM configuration in nonmagnetic crystals. By analyzing all 230 crystallographic SGs under the constraints of topological charge conservation and symmetry-enforced Weyl-point orbits, we identify 12 and 23 candidate SGs capable of hosting this configuration without and with SOC, respectively. We further predict two boron allotropes, TBR-B$_{48}$ and chiral OBD-B$_{48}$, as material realizations of this charge-compensated Weyl state. The two systems exhibit hourglass-type and non-hourglass-type C-3 WPs, respectively, demonstrating distinct realizations of the same global Weyl configuration. Surface-state calculations reveal sixfold and threefold Fermi-arc structures associated with the projections of the C-3 and C-1 Weyl sectors, while the opposite structural chiralities of OBD-B$_{48}$ reverse all Weyl charges. Finally, strain engineering in TBR-B$_{48}$ demonstrates the symmetry-controlled transition from C-3 WPs to multiple C-1 WPs through the breaking of the protecting $C_6$ rotational symmetry. Our work establishes a symmetry-based framework for understanding and designing nonmagnetic WSMs with charge-three and charge-one Weyl-point coexistence.

\section{Computational Methods}\label{sec:methods}
First-principles calculations were performed within density functional theory (DFT) using the Vienna $ab$ initio Simulation Package (VASP)~\cite{VASP}. The projector augmented-wave (PAW) method was employed to describe the ion--electron interactions~\cite{PAW}, and the exchange--correlation functional was treated within the Perdew--Burke--Ernzerhof (PBE) generalized gradient approximation (GGA)~\cite{PBE}. A plane-wave cutoff energy of 520 eV was used. The Brillouin-zone integrations were performed using Monkhorst--Pack~\cite{MonkhorstPack} $k$-point meshes of $6\times6\times8$ and $6\times6\times4$ for TBR-B$_{48}$ and chiral OBD-B$_{48}$, respectively. The dynamical stability was evaluated using phonon calculations implemented in PHONOPY~\cite{Phonopy,Togo2008} package with $2\times2\times2$ supercells. The mechanical stability was assessed from elastic constants obtained through the energy--strain method~\cite{Elastic,ElasticStability}. For topological analysis, maximally localized Wannier functions~\cite{Wannier90Updated} were constructed to obtain tight-binding (TB) Hamiltonians. These Hamiltonians were then used in WannierTools~\cite{WannierTools} to calculate surface spectra via the iterative Green's-function method, locate WPs throughout the Brillouin zone, and determine their topological charges from the evolution of Wannier charge centers on enclosing surfaces.

\begin{table*}[p]
\caption{Complete symmetry-pinned high-symmetry C-1/C-3 candidate channels retained for the $2\times\{3,-1,-1,-1\}$-IWM configuration after local-charge, magnetic-star, and fixed-gap-parity screening. The star multiplicity $m$ is defined by $m=[G:G_{\mathbf{k}}]$. The symbol $n$ denotes the band index of the highest occupied state. PG denotes the full crystallographic point
group.  Parentheses denote crossings of distinct one-dimensional branches;
braces and single labels denote essential corepresentations (coreps).  HSP/HSL means
high-symmetry point/line.  Rows sharing the same physical channel are
consolidated, with every SG--locus and SG--representation mapping retained
explicitly.  For SG 155, $(a)F$ and $(b)F$ are two labels of the same star,
with generators $C_{21}^{\prime},\mathcal{T}$ and
$C_{23}^{\prime},\mathcal{T}$, respectively.  The IRR notation follows
Ref.~\cite{BradleyCracknell}.}
\label{tab:spacegroups}
\centering
\begingroup
\setlength{\tabcolsep}{1.2pt}
\renewcommand{\arraystretch}{1.25}
\setlength{\extrarowheight}{0.8mm}
\scriptsize
\begin{tabular}{p{0.060\textwidth}p{0.258\textwidth}p{0.035\textwidth}p{0.020\textwidth}>{\centering\arraybackslash}p{0.190\textwidth}p{0.389\textwidth}}
	\toprule
	\textbf{Species} & \textbf{SGs (HSP/HSL)} & \textbf{PG} & \textbf{$m$} &
	\tcellc{0.185\textwidth}{\textbf{Little-group}\\\textbf{generators}} & \textbf{IRRs/coreps} \\
	\midrule
	\multicolumn{6}{c}{\textbf{Without SOC: single-valued representations; $\Gamma$A C-3 pair}} \\
	\midrule
	C-3 WP & 168--173 ($\Gamma$A) & $C_6$ & 2 & $C_6^{+}$ & $(R_1,R_4)$; $(R_2,R_5)$; $(R_3,R_6)$ \\
	& 177--182 ($\Gamma$A) & $D_6$ & 2 & $C_6^{+},C_{23}^{\prime\prime}\mathcal{T}$ & $(R_1,R_4)$; $(R_2,R_5)$; $(R_3,R_6)$ \\
	C-1 WP & 168--173 ($\Gamma$A) & $C_6$ & 2 & $C_6^{+}$ & $(R_1,R_2)$; $(R_1,R_6)$; $(R_2,R_3)$; $(R_3,R_4)$; $(R_4,R_5)$; $(R_5,R_6)$ \\
	& 168--173 (KH) & $C_6$ & 4 & $C_3^{+}$ & $(R_1,R_2)$; $(R_1,R_3)$; $(R_2,R_3)$ \\
	& 168--173 (ML) & $C_6$ & 6 & $C_2$ & $(R_1,R_2)$ \\
	& 168, 171 (H,K);\newline 169, 170, 172, 173 (K) & $C_6$ & 2 & $C_3^{+},C_6^{+}\mathcal{T}$ & $\{R_2,R_3\}$ \\
	& 172 (H) & $C_6$ & 2 & $C_3^{-},C_6^{+}\mathcal{T}$ & $\{R_2,R_3\}$ \\
	& 177--182 ($\Gamma$A) & $D_6$ & 2 & $C_6^{+},C_{23}^{\prime\prime}\mathcal{T}$ & $(R_1,R_2)$; $(R_1,R_6)$; $(R_2,R_3)$; $(R_3,R_4)$; $(R_4,R_5)$; $(R_5,R_6)$ \\
	& 177--182 (KH) & $D_6$ & 4 & $C_3^{+},C_{22}^{\prime}\mathcal{T}$ & $(R_1,R_2)$; $(R_1,R_3)$; $(R_2,R_3)$ \\
	& 177, 180, 181 (AH) & $D_6$ & 6 & $C_{22}^{\prime\prime},C_{22}^{\prime}\mathcal{T}$ & 177, 181: $(R_1,R_2)$; 180: $(R_2,R_5)$ \\
	& 177, 180, 181 (AL,$\Gamma$M);\newline 178, 179, 182 ($\Gamma$M) & $D_6$ & 6 & $C_{21}^{\prime},C_{21}^{\prime\prime}\mathcal{T}$ & $(R_1,R_2)$ \\
	& 177, 180, 181 (LH,MK);\newline 178, 179, 182 (MK) & $D_6$ & 6 & $C_{21}^{\prime\prime},C_2\mathcal{T}$ & $(R_1,R_2)$ \\
	& 177--182 (ML) & $D_6$ & 6 & $C_2,C_{21}^{\prime\prime}\mathcal{T}$ & $(R_1,R_2)$ \\
	& 177--182 ($\Gamma$K) & $D_6$ & 6 & $C_{22}^{\prime\prime},C_{22}^{\prime}\mathcal{T}$ & $(R_1,R_2)$ \\
	& 177, 180 (H,K);\newline 178, 179, 182 (K) & $D_6$ & 2 & $C_3^{+},C_{21}^{\prime\prime},C_6^{+}\mathcal{T}$ & $R_3$ \\
	& 181 (H,K) & $D_6$ & 2 & $C_3^{-},C_{21}^{\prime\prime},C_6^{+}\mathcal{T}$ & $R_3$ \\
	\midrule
	\multicolumn{6}{c}{\textbf{With SOC, odd $n$: eight TRIM nodes; C-1 stars $3+3$}} \\
	\midrule
	C-3 WP & 143, 145 (A,$\Gamma$);\newline 144 ($\Gamma$);\newline 146 (Z,$\Gamma$) & $C_3$ & 1 & $C_3^{+},\mathcal{T}$ & $\{R_4,R_4\}$ \\
	& 144 (A) & $C_3$ & 1 & $C_3^{-},\mathcal{T}$ & $\{R_4,R_4\}$ \\
	& 149, 153 (A,$\Gamma$);\newline 151 ($\Gamma$);\newline 155 (Z,$\Gamma$) & $D_3$ & 1 & $C_3^{+},C_{21}^{\prime},\mathcal{T}$ & $\{R_3,R_4\}$ \\
	& 151 (A) & $D_3$ & 1 & $C_3^{-},C_{21}^{\prime},\mathcal{T}$ & $\{R_3,R_4\}$ \\
	& 150, 154 (A,$\Gamma$);\newline 152 ($\Gamma$) & $D_3$ & 1 & $C_3^{+},C_{21}^{\prime\prime},\mathcal{T}$ & $\{R_3,R_4\}$ \\
	& 152 (A) & $D_3$ & 1 & $C_3^{-},C_{21}^{\prime\prime},\mathcal{T}$ & $\{R_3,R_4\}$ \\
	& 168, 171, 172 (A,$\Gamma$) & $C_6$ & 1 & $C_6^{+},\mathcal{T}$ & $\{R_4,R_{10}\}$ \\
	& 177, 180, 181 (A,$\Gamma$) & $D_6$ & 1 & $C_6^{+},C_{21}^{\prime},\mathcal{T}$ & $R_7$ \\
	C-1 WP & 143--145 (M,L);\newline 146 (F,L) & $C_3$ & 3 & $\mathcal{T}$ & $\{R_2,R_2\}$ \\
	& 155 ($(a)F$) & $D_3$ & 3 & $C_{21}^{\prime},\mathcal{T}$ & $\{R_2,R_4\}$ \\
	& 155 ($(b)F$) & $D_3$ & 3 & $C_{23}^{\prime},\mathcal{T}$ & $\{R_2,R_4\}$ \\
	& 155 (L) & $D_3$ & 3 & $C_{22}^{\prime},\mathcal{T}$ & $\{R_2,R_4\}$ \\
	& 149, 151, 153 (L,M) & $D_3$ & 3 & $C_{21}^{\prime},\mathcal{T}$ & $\{R_2,R_4\}$ \\
	& 150, 152, 154 (L,M) & $D_3$ & 3 & $C_{21}^{\prime\prime},\mathcal{T}$ & $\{R_2,R_4\}$ \\
	& 168, 171, 172 (L,M) & $C_6$ & 3 & $C_2,\mathcal{T}$ & $\{R_2,R_4\}$ \\
	& 177, 180, 181 (L,M) & $D_6$ & 3 & $C_2,C_{21}^{\prime\prime},\mathcal{T}$ & $R_5$ \\
	\bottomrule
\end{tabular}
\endgroup
\end{table*}

\begin{table*}[t!]
\centering
\begingroup
\setlength{\tabcolsep}{1.2pt}
\renewcommand{\arraystretch}{1.25}
\setlength{\extrarowheight}{0.8mm}
\scriptsize
\begin{tabular}{p{0.060\textwidth}p{0.258\textwidth}p{0.035\textwidth}p{0.020\textwidth}>{\centering\arraybackslash}p{0.190\textwidth}p{0.389\textwidth}}
\multicolumn{6}{c}{\normalsize\textbf{TABLE I. (Continued)}} \\[2pt]
\toprule
\textbf{Species} & \textbf{SGs (HSP/HSL)} & \textbf{PG} & \textbf{$m$} &
\tcellc{0.185\textwidth}{\textbf{Little-group}\\\textbf{generators}} & \textbf{IRRs/coreps} \\
\midrule
\multicolumn{6}{c}{\textbf{With SOC, even $n$: non-TRIM nodes; C-3 star 2}} \\
\midrule
C-3 WP & 168--173 ($\Gamma$A) & $C_6$ & 2 & $C_6^{+}$ & $(R_2,R_8)$; $(R_4,R_{10})$; $(R_6,R_{12})$ \\
& 177--182 ($\Gamma$A) & $D_6$ & 2 & $C_6^{+},C_{23}^{\prime\prime}\mathcal{T}$ & $(R_2,R_8)$; $(R_4,R_{10})$; $(R_6,R_{12})$ \\
C-1 WP & 168--173 ($\Gamma$A) & $C_6$ & 2 & $C_6^{+}$ & $(R_2,R_{12})$; $(R_2,R_4)$; $(R_4,R_6)$; $(R_6,R_8)$; $(R_8,R_{10})$; $(R_{10},R_{12})$ \\
& 169--173 (KH) & $C_6$ & 4 & $C_3^{+}$ & $(R_2,R_4)$; $(R_2,R_6)$; $(R_4,R_6)$ \\
& 169, 170, 173 (ML) & $C_6$ & 6 & $C_2$ & $(R_2,R_4)$ \\
& 171 (H,K);\newline 169, 170, 172, 173 (K) & $C_6$ & 2 & $C_3^{+},C_6^{+}\mathcal{T}$ & $\{R_2,R_6\}$ \\
& 172 (H) & $C_6$ & 2 & $C_3^{-},C_6^{+}\mathcal{T}$ & $\{R_2,R_6\}$ \\
& 177--182 ($\Gamma$A) & $D_6$ & 2 & $C_6^{+},C_{23}^{\prime\prime}\mathcal{T}$ & $(R_2,R_{12})$; $(R_2,R_4)$; $(R_4,R_6)$; $(R_6,R_8)$; $(R_8,R_{10})$; $(R_{10},R_{12})$ \\
& 178--182 (KH) & $D_6$ & 4 & $C_3^{+},C_{22}^{\prime}\mathcal{T}$ & $(R_2,R_4)$; $(R_2,R_6)$; $(R_4,R_6)$ \\
& 178, 179, 182 (MK) & $D_6$ & 6 & $C_{21}^{\prime\prime},C_2\mathcal{T}$ & $(R_2,R_4)$ \\
& 178, 179, 182 (ML) & $D_6$ & 6 & $C_2,C_{21}^{\prime\prime}\mathcal{T}$ & $(R_2,R_4)$ \\
& 178, 179, 182 ($\Gamma$K) & $D_6$ & 6 & $C_{22}^{\prime\prime},C_{22}^{\prime}\mathcal{T}$ & $(R_2,R_4)$ \\
& 180 (H,K);\newline 178, 179, 182 (K) & $D_6$ & 2 & $C_3^{+},C_{21}^{\prime\prime},C_6^{+}\mathcal{T}$ & $R_6$ \\
& 181 (H,K) & $D_6$ & 2 & $C_3^{-},C_{21}^{\prime\prime},C_6^{+}\mathcal{T}$ & $R_6$ \\
\bottomrule
\end{tabular}

\endgroup
\end{table*}

\section{Results}
\label{sec:results}

\subsection{Symmetry constraints for the $2\times\{3,-1,-1,-1\}$-IWM configuration}

We first establish the symmetry requirements for realizing the target Weyl-node complex. We consider a nonmagnetic Weyl semimetal phase in which all band crossings within a fixed direct band gap are isolated twofold WPs. The target topological-charge configuration is defined as

\begin{equation}
	\mathcal{C}_{31}
	=
	\{3s,3s,\underbrace{-s,\ldots,-s}_{6}\},
	\qquad s=\pm1 ,
\end{equation}

where the two choices of $s$ correspond to an overall reversal of chirality. This configuration consists of two identical irreducible Weyl molecules (IWMs)~\cite{IWM}, $2\times\{3,-1,-1,-1\}$, where the notation represents multiset repetition rather than independent band sectors or charge multiplication.

The realization of this configuration is constrained by both local crystalline symmetry and global topological charge conservation. The Nielsen--Ninomiya theorem requires the total chiral charge in the Brillouin zone to vanish~\cite{NielsenNinomiya1,NielsenNinomiya2}. In time-reversal-symmetric systems, a WP at $\mathbf{k}$ and its time-reversal partner at $-\mathbf{k}$ carry the same topological charge,

\begin{equation}
	C(-\mathbf{k})=C(\mathbf{k}).
\end{equation}

Therefore, a time-reversal-related pair of C-3 WPs contributes a net charge of $\pm6$ and must be compensated by six C-1 WPs carrying opposite chirality. Consequently, the coexistence of C-3 and C-1 WPs cannot be determined solely from the local existence of individual high-charge nodes, but requires a symmetry-compatible global distribution of all WPs.

For a crystallographic SG $G$, we impose six symmetry constraints to identify candidate Weyl-node complexes.

\textit{(i) Intrinsic local availability.}
The SG must simultaneously allow both C-3 and C-1 WPs through symmetry-protected high-symmetry channels~\cite{Encyclopedia} or symmetry-compatible Weyl-node configurations. This condition provides only a necessary local criterion and does not guarantee the existence of the complete eight-node configuration.

\textit{(ii) Charge-pattern compatibility.}
For a symmetry operation $a=\{R_a|\boldsymbol{\tau}_a\}\mathcal{T}^{\epsilon_a}$, the Weyl charge transforms as

\begin{equation}
	C(a\mathbf{k})=\det(R_a)C(\mathbf{k}).
\end{equation}

Since all WPs in the target configuration have fixed chirality signs, symmetry operations relating nodes within the same orbit must preserve their charges. Therefore, improper operations with $\det(R_a)=-1$ are incompatible with the target configuration because they generate additional WPs with opposite chirality. This requirement restricts the candidate groups to chiral Sohncke SGs.

\textit{(iii) Charge-resolved Weyl-star constraint.}
The multiplicity of symmetry-related WPs is determined by the size of the momentum star,

\begin{equation}
	m(\mathbf{k})
	=
	[G:G_{\mathbf{k}}],
\end{equation}

where $G_{\mathbf{k}}$ is the little group of $\mathbf{k}$. The symmetry-generated Weyl-point orbits must exactly satisfy

\begin{equation}
	\sum_{\alpha\in\mathcal{I}_{3}}m_{\alpha}=2,
	\qquad
	\sum_{\alpha\in\mathcal{I}_{1}}m_{\alpha}=6 ,
\end{equation}

where $\mathcal{I}_{3}$ and $\mathcal{I}_{1}$ denote the selected C-3 and C-1 Weyl sectors, respectively. Thus, the C-3 sector must form either a two-arm star or two symmetry-inequivalent one-arm stars, whereas the C-1 sector must contain closed symmetry orbits whose total multiplicity is exactly six.

\textit{(iv) Representation-resolved node selection.}
The selected WPs must correspond simultaneously to the correct SG, momentum position, little-group irreducible (co)representation (IRR), and Weyl-particle type. In particular, an essential two-dimensional corepresentation and an accidental crossing between two one-dimensional representations represent different mechanisms and must be treated separately. The representation data determine the allowed band coupling and therefore the possible Weyl charge.

\textit{(v) Exclusion of extended degeneracies.}
For spinful systems, time-reversal symmetry enforces Kramers degeneracy at time-reversal-invariant momenta (TRIMs),

\begin{equation}
	E_{2j-1}(\mathbf{T}) =	E_{2j}(\mathbf{T}).
\end{equation}

Therefore, candidate configurations must be examined separately according to the band-index ($n$) parity. For odd-index gaps, TRIM crossings may directly contribute to the required eight WPs, whereas for even-index gaps, TRIM degeneracies involving multiple Kramers pairs must remain outside the selected gap. In addition, screw-related antiunitary symmetries may enforce nodal planes, which must be excluded because the present work considers isolated WPs only.

\textit{(vi) Global compatibility condition.}
Finally, the selected local representations and Weyl orbits must form a globally compatible band connectivity. The compatibility relations~\cite{MagneticTQC} between high-symmetry points (HSPs) and lines (HSLs) must preserve the required band ordering and avoid additional WPs, nodal lines, nodal surfaces, or higher-fold degeneracies within the same direct gap. Therefore, satisfying local representation conditions alone is insufficient; a complete realization requires simultaneous satisfaction of symmetry, topology, and band-connectivity constraints.

Applying these six criteria to all 230 crystallographic SGs, using the symmetry database of emergent particles in three-dimensional crystals~\cite{Encyclopedia}, yields the candidate SGs capable of hosting the $2\times\{3,-1,-1,-1\}$-IWM configuration. It should be noted that the symmetry classification summarized in Table~\ref{tab:spacegroups} focuses on WPs protected by crystalline symmetries at HSPs and HSLs, where little-group representations can be explicitly assigned. General-momentum WPs that are not individually pinned by symmetry, such as the compensating C-1 WPs in TBR-B$_{48}$, are not listed as separate symmetry channels in Table~\ref{tab:spacegroups}. Nevertheless, their existence is fully incorporated into the six symmetry-screening criteria above through the charge-resolved Weyl-star constraint, representation compatibility, and global band-connectivity requirements. Therefore, the classification provides the symmetry framework for constructing the complete double $\{3,-1,-1,-1\}$-IWM configuration rather than a catalogue restricted exclusively to symmetry-pinned C-1 WPs. Without SOC, only the hexagonal SGs 168\text{--}173, 177\text{--}182, satisfy the required conditions, giving 12 candidate groups. When SOC is included, double-valued representations lead to additional allowed symmetry channels, resulting in 23 candidate SGs: 143\text{--}146, 149\text{--}155, 168\text{--}173, 177\text{--}182. The complete momentum locations, symmetry generators, and irreducible (co)representations of the C-3 and C-1 Weyl sectors are summarized in Table~\ref{tab:spacegroups}.

\subsection{Low-energy effective model of C-3 WPs}

To understand the local topological origin of the C-3 WPs, we construct a symmetry-constrained two-band $k\cdot p$ Hamiltonian~\cite{kdotp,zhangkp} near the band-crossing point. In the basis of the two crossing states, the effective Hamiltonian is expressed as

\begin{equation}
	H(\mathbf{k})
	=
	M(\mathbf{k})\sigma_z
	+
	F(\mathbf{k})\sigma_+
	+
	F^*(\mathbf{k})\sigma_- ,
\end{equation}

where $\sigma_{\pm}=(\sigma_x\pm i\sigma_y)/2$.

For a C-3 WP protected by a sixfold rotational symmetry, the two crossing states possess opposite rotation eigenvalues~\cite{multiweyl},

\begin{equation}
	\frac{\lambda_1}{\lambda_2}=-1 .
\end{equation}

The rotational selection rule forbids lower-order in-plane hybridization terms, and the leading allowed off-diagonal coupling becomes

\begin{equation}
	F(\mathbf{k})
	=
	m_1k_+^3+m_2k_-^3 ,
\end{equation}

with $k_{\pm}=k_x\pm ik_y$.

The diagonal term is expanded as

\begin{equation}
	M(\mathbf{k})
	=
	n_1k_\perp^2+n_2k_z+\cdots .
\end{equation}

Therefore, the effective Hamiltonian near the C-3 WP takes the form

\begin{equation}
	H(\mathbf{k})
	=
	\begin{pmatrix}
		M(\mathbf{k})&
		m_1k_+^3+m_2k_-^3
		\\
		(m_1k_+^3+m_2k_-^3)^*&
		-M(\mathbf{k})
	\end{pmatrix}.
\end{equation}

The cubic winding of the interband coupling gives rise to a Berry monopole carrying three units of topological charge~\cite{multiweyl,Encyclopedia}, $|C|=3$, which is the characteristic topological feature of the C-3 WPs realized in TBR-B$_{48}$ and OBD-B$_{48}$ [see Fig. S1 of the Supplemental Material (SM)~\cite{SM54}].

\begin{figure*}[t!]
\centering
\includegraphics[width=1.0\textwidth]{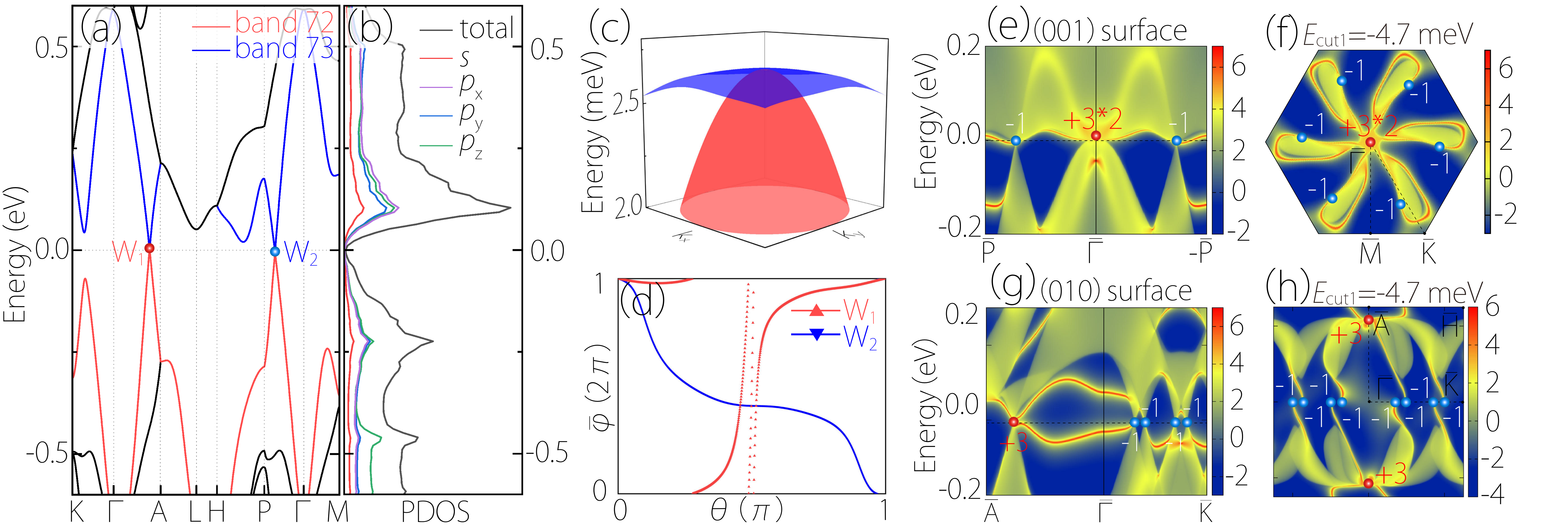}
\caption{
Electronic structure and topological properties of TBR-B$_{48}$. (a) Bulk band structures. (b) projected density of states. The bands near the Fermi level are clean and the crossing along the $\Gamma$-$A$ line forms the C-3 Weyl pair. (c) Three-dimensional band dispersion around the C-3 WP, showing the characteristic cubic Weyl dispersion. (d) Evolution of Wannier charge centers on enclosing spheres, confirming the topological charges of $C=+3$ for the W$_1$ WPs and $C=-1$ for the compensating W$_2$ WPs. (e) Surface spectral function on the (001) surface and (f) corresponding constant-energy contour showing six Fermi-arc branches connecting the co-projected C-3 Weyl pair to the C-1 projections. (g) Surface spectrum on the (010) surface and (h) constant-energy contour revealing three Fermi arcs associated with each separated C-3 projection. These results establish TBR-B$_{48}$ as a realization of the $2\times\{3,-1,-1,-1\}$-IWM configuration.
}
\label{fig:tbr}
\end{figure*}

\begin{table*}[t!]
\makeatletter
\long\def\@makecaption#1#2{%
  \par
  \vskip\abovecaptionskip
  \begingroup
    \small\rmfamily
    \flushing
    \@make@capt@title{#1}{#2}\par
  \endgroup
  \vskip\belowcaptionskip
}
\makeatother
\caption{Energies, coordinates, topological charges, and multiplicities of the C-1 WP and C-3 WP in TBR-B$_{48}$, \emph{l}-OBD-B$_{48}$, and \emph{r}-OBD-B$_{48}$.}
\label{tab:weylpoints}
\centering
\begingroup
\small
\setlength{\tabcolsep}{6.5pt}
\renewcommand{\arraystretch}{1.12}
\begin{tabular*}{\textwidth}{@{\extracolsep{\fill}}lccccc@{}}
\toprule
Structure & WP & $E$ (meV) & Coordinate $(k_1,k_2,k_3)$ & charge & multiplicity \\
\midrule
\multirow{5}{*}{TBR-B$_{48}$}
& W$_1$ & 2.49 & $(0,0,\pm0.381)$ & $+3$ &  2  \\
& W$_2$ & -4.70 & $(\pm0.463,\mp0.264,0)$ & $-1$ & \multirow{3}{*}{6} \\
& W$_2$ & -4.70 & $\pm(-0.264,-0.199,0)$ & $-1$ & \\
& W$_2$ & -4.70 & $(\pm0.199,\mp0.463,0)$ & $-1$ & \\
\midrule
\multirow{5}{*}{\emph{l}-OBD-B$_{48}$}
& W$_1$ & -0.103 & $(0,0,\pm0.104)$ & $-3$ & 2 \\
& W$_2$ & 0.005 & $(0,\pm0.5,\pm0.314)$ & $+1$ & \multirow{4}{*}{6} \\
& W$_2$ & 0.005 & $(\pm0.5,0,\pm0.314)$ & $+1$ & \\
& W$_2$ & 0.005 & $(+0.5,-0.5,\pm0.314)$ & $+1$ & \\
& W$_2$ & 0.005 & $(-0.5,+0.5,\pm0.314)$ & $+1$ & \\
\midrule
\multirow{5}{*}{\emph{r}-OBD-B$_{48}$}
& W$_1$ & -0.105 & $(0,0,\pm0.104)$ & $+3$ & 2 \\
& W$_2$ & 0.002 & $(0,\pm0.5,\pm0.314)$ & $-1$ & \multirow{4}{*}{6} \\
& W$_2$ & 0.002 & $(\pm0.5,0,\pm0.314)$ & $-1$ & \\
& W$_2$ & 0.002 & $(+0.5,-0.5,\pm0.314)$ & $-1$ & \\
& W$_2$ & 0.002 & $(-0.5,+0.5,\pm0.314)$ & $-1$ & \\
\bottomrule
\end{tabular*}
\endgroup
\end{table*}

\subsection{Crystal structures and stability}\label{sec:structure}

The symmetry classification establishes the crystallographic conditions required for realizing the
$2\times\{3,-1,-1,-1\}$-IWM configuration. Guided by these criteria and the candidate SGs, we performed a targeted search for boron allotropes using the RG2 structure-search code~\cite{RG2}, exploiting the intrinsically weak spin-orbit coupling (SOC) of boron, which allows high-Chern-number band crossings to remain primarily governed by crystalline symmetry~\cite{Charge2WeylDirac,OddNodeWeylDirac,ChargeTwoTriplePoint,FourDoubleWeyl}. Two three-dimensional boron allotropes, a triangular-boron-ribbon structure (TBR-B$_{48}$) and an octahedral boron dimer (OBD-B$_{48}$) structure with intrinsic chirality, are identified as representative realizations belonging to the candidate SGs obtained from the symmetry screening.

TBR-B$_{48}$ crystallizes in the chiral hexagonal SG $P6_3$ (No.~173), which is one of the symmetry-allowed groups supporting the target Weyl-node complex without SOC [see Table~\ref{tab:spacegroups}]. The optimized structure contains 48 boron atoms in the primitive cell. As shown in Fig.~\ref{fig:structure}(a), the $2\times2\times1$ supercell features a three-dimensional boron network composed of buckled triangular boron nanoribbons (B$_1$, Fig.~\ref{fig:structure}(b)), triangular boron units (B$_3$, shown in gray), and double-bilayer triangular boron units (B$_2$, Fig.~\ref{fig:structure}(c)). The basic structural motif enclosed by the green dashed region consists of B$_1$ atoms forming a buckled nanoribbon with width $n=5$ [Fig.~\ref{fig:structure}(b)]. Adjacent nanoribbons are connected by triangular boron units composed of gray B$_3$ atoms. Figure~\ref{fig:structure}(c) shows a side view of the structural unit enclosed by the black dashed box, in which two in-plane triangular boron units are rotated by $60^\circ$ relative to each other along the $c$ axis and stacked to form a double-bilayer triangular boron structure. After full structural relaxation, the optimized lattice parameters are $a=b=8.47$~\AA{} and $c=5.55$~\AA{}, with bond angles ranging from $50.18^\circ$ to $166.73^\circ$. Detailed structural information is provided in Table S1 of the SM~\cite{SM54}.

The other candidate material, OBD-B$_{48}$, exhibits two enantiomorphic structures, namely left-handed \emph{l}-OBD-B$_{48}$ and right-handed \emph{r}-OBD-B$_{48}$. The two structures crystallize in the chiral SGs $P6_2$ (No.~171) and $P6_4$ (No.~172), respectively, which are also included in the symmetry-allowed space-group list. As shown in Figs.~\ref{fig:structure}(f,g), both structures contain 48 boron atoms in the primitive cell and exhibit intrinsic structural chirality. The framework is composed of three characteristic structural motifs: buckled boron nanotubes formed by stacked eight-membered rings [Figs.~\ref{fig:structure}(h,j,k)], single-atom-wide boron helical chains [Figs.~\ref{fig:structure}(l,m)], and bridging boron atoms (gray) connecting these structural units. The optimized lattice parameters are $a=b=7.66$~\AA{} and $c=7.51$~\AA{}, with bond angles ranging from $50.39^\circ$ to $172.45^\circ$.

The structural stability of these two boron allotropes was systematically evaluated. The calculated elastic constants (Table SII of the SM~\cite{SM54}) satisfy the corresponding Born mechanical-stability criteria~\cite{ElasticStability}, namely, $C_{11} > |C_{12}|$, $2C_{13}^2 < C_{33}(C_{11} + C_{12})$, $C_{44} > 0$, and $C_{66} > 0$, thereby confirming their mechanical stability. Furthermore, the calculated phonon spectra show no imaginary frequencies throughout the BZ (Fig. S2 in the SM~\cite{SM54}), demonstrating their dynamical stability. Therefore, TBR-B$_{48}$ and OBD-B$_{48}$ provide structurally stable platforms for investigating symmetry-constrained coexistence of C-3 and C-1 WPs.

\subsection{Electronic structures and topological properties of TBR-B$_{48}$}
\label{sec:tbr}

We first investigate the electronic and topological properties of TBR-B$_{48}$, whose SG $P6_3$ (No.~173) belongs to the symmetry class supporting the
$2\times\{3,-1,-1,-1\}$-IWM configuration. Because boron is a light element and the SOC effect is negligible, the electronic structures discussed below are calculated without SOC, such that the high-Chern-number Weyl crossings are governed primarily by crystalline symmetry~\cite{multiweyl,Encyclopedia}.

Figure~\ref{fig:tbr}(a) shows the bulk band structure of TBR-B$_{48}$ along the high-symmetry paths. The bands near the Fermi level are remarkably clean, providing a favorable environment for identifying isolated Weyl crossings. The 72nd and 73rd bands cross along the $\Gamma$-$A$ line, generating a pair of WPs labeled W$_1$, which correspond to the neck points of the hourglass-shaped band structure. These two nodes are located at $E-E_F=2.49~\mathrm{meV}$, and are related by time-reversal symmetry. The crossing states belong to the two-dimensional irreducible representation
$\Delta_3+\Delta_4$ of the little group along the $\Gamma$-$A$ line, consistent with the symmetry requirement for a C-3 WP derived from the effective model.

To identify all WPs associated with the global gap between the two bands, we perform a full Brillouin-zone search for crossings between the lowest conduction band (72nd) and the highest valence band (73rd). In addition to the two W$_1$ nodes located along $\Gamma$-$A$, six additional WPs, denoted as W$_2$, are found in the $k_z=0$ plane [Fig.~\ref{fig:structure}(e)]. The energies, momenta, multiplicities, and topological charges of all WPs are summarized in Table~\ref{tab:weylpoints}. The two W$_1$ nodes form the C-3 sector with multiplicity two, whereas the six W$_2$ nodes constitute the compensating C-1 sector, exactly satisfying the charge-count requirement of the $2\times\{3,-1,-1,-1\}$-IWM configuration.

The projected density of states shown in Fig.~\ref{fig:tbr}(b) exhibits a pronounced minimum around the Fermi level, confirming the semimetallic character of TBR-B$_{48}$. The electronic states near the Weyl crossings are mainly contributed by strongly hybridized boron $p$ orbitals, indicating that the high-charge topology originates from the intrinsic bonding network of the boron framework rather than from heavy-element SOC effects.

To further characterize the local dispersion of the WPs, we calculate the three-dimensional band dispersion around W$_1$ [Fig.~\ref{fig:tbr}(c)]. The crossing exhibits the cubic dispersion expected for a C-3 WP, with a linear dispersion component along the rotational axis [Fig.~\ref{fig:tbr}(a)] and higher-order momentum dependence in the transverse directions. The corresponding dispersion of the compensating W$_2$ nodes is shown in Fig.~S3(a) of the SM~\cite{SM54}. Unlike W$_1$, the W$_2$ nodes exhibit conventional linear Weyl dispersion, consistent with their unit topological charge.

The topological charges of the WPs are determined from the evolution of Wannier charge centers on enclosing spherical surfaces. As shown in Fig.~\ref{fig:tbr}(d), the Wannier charge-center flow of W$_1$ exhibits a winding number of three, confirming $C(\mathrm{W}_1)=+3$. In contrast, the six W$_2$ nodes each carry $C(\mathrm{W}_2)=-1$. Therefore, the total chiral charge in the BZ satisfies $2\times(+3)+6\times(-1)=0$, consistent with the constraint imposed by the Nielsen--Ninomiya no-go theorem and confirming that TBR-B$_{48}$ realizes the predicted charge-compensated $2\times\{3,-1,-1,-1\}$-IWM complex.

The bulk topology is further verified through surface-state calculations. Because the two time-reversal-related W$_1$ nodes possess the same chirality, their surface projections determine the connectivity of the Fermi arcs. On the (001) surface, the two W$_1$ nodes are projected onto the same surface momentum point $\bar{\Gamma}$ [Fig.~\ref{fig:structure}(e) and Figs.~\ref{fig:tbr}(e,f)], resulting in an accumulated projected charge of $C_{\mathrm{proj}}=+6$. Accordingly, the constant-energy contour in Fig.~\ref{fig:tbr}(f) reveals six Fermi-arc branches extending from the co-projected C-3 Weyl sector toward the six projected C-1 WPs, forming a sixfold helicoid Fermi-arc structure.

By contrast, on the (010) surface, the two C-3 WPs project onto different surface momenta. As shown in Fig.~\ref{fig:tbr}(g), each projected C-3 node connects to three Fermi arcs terminating at the projected C-1 nodes. The corresponding constant-energy contour in Fig.~\ref{fig:tbr}(h) displays the expected threefold Fermi-arc connectivity. The different surface manifestations originate solely from the projection geometry of the bulk Weyl-node complex, demonstrating the direct correspondence between bulk charge distribution and surface topology.

Together, the bulk band structure, irreducible representations, Wannier charge-center evolution, and surface-state calculations establish TBR-B$_{48}$ as a concrete realization of the symmetry-constrained $2\times\{3,-1,-1,-1\}$-IWM Weyl complex.

\begin{figure*}[t!]
\centering
\includegraphics[width=1.0\textwidth]{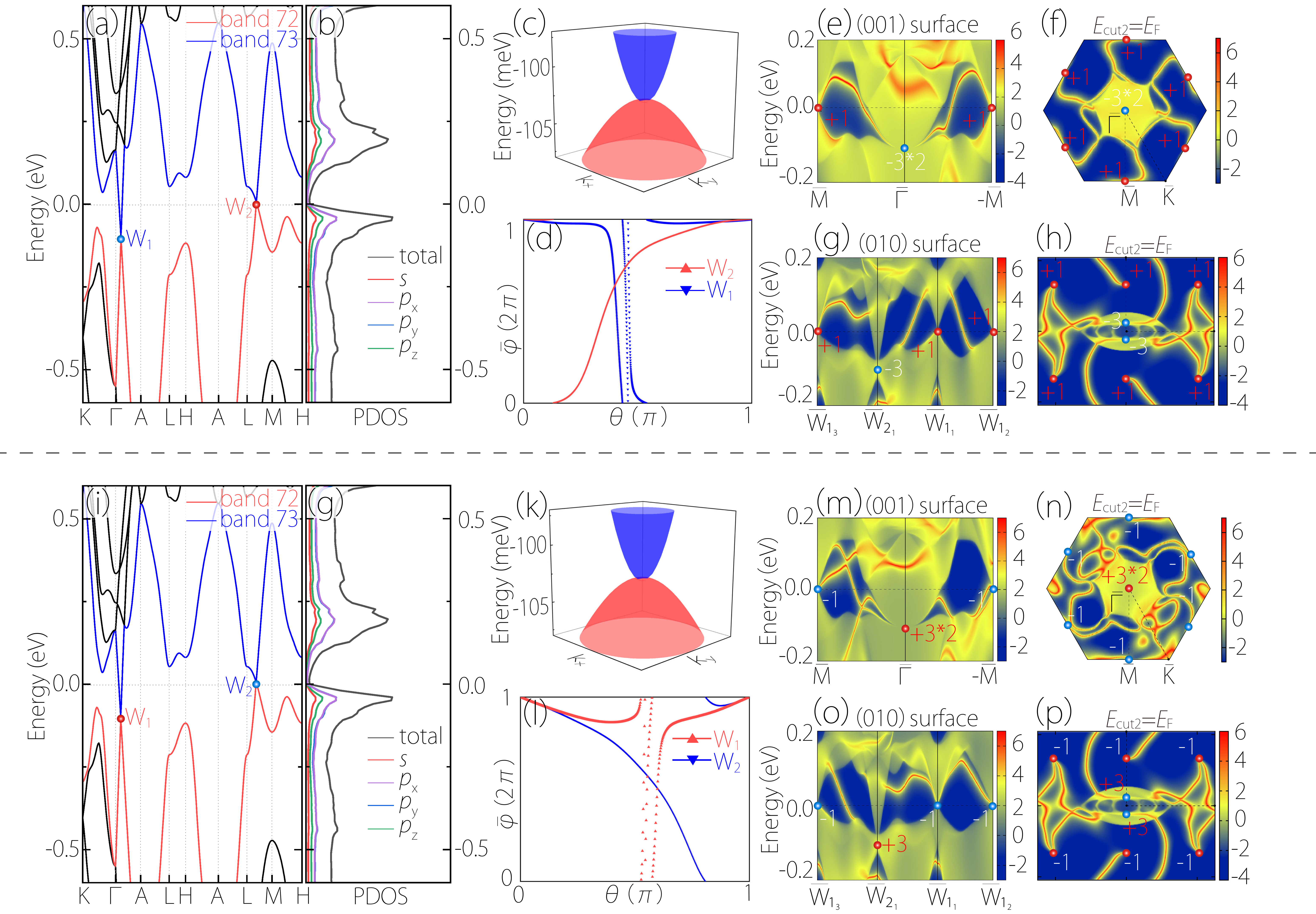}
\caption{Chirality-dependent Weyl topology in chiral OBD-B$_{48}$. (a,b) Bulk electronic structures and projected density of states of left-handed OBD-B$_{48}$. (c) Three-dimensional dispersion around the C-3 WP. (d) Wannier charge-center evolution confirming $C=-3$ for the W$_1$ WPs and $C=+1$ for the W$_2$ WPs. (e,f) Surface states and constant-energy contour on the (001) surface showing sixfold Fermi-arc connectivity. (g,h) Surface states on the (010) surface exhibiting three Fermi arcs from each projected C-3 WP. (i--p) Corresponding results for right-handed OBD-B$_{48}$. The reversal of structural chirality reverses the signs of all Weyl charges while preserving the number and topology of the surface Fermi arcs, demonstrating the direct correspondence between crystal chirality and momentum-space chirality.
}
\label{fig:obd}
\end{figure*}

\begin{figure*}[t!]
\centering
\includegraphics[width=0.98\textwidth]{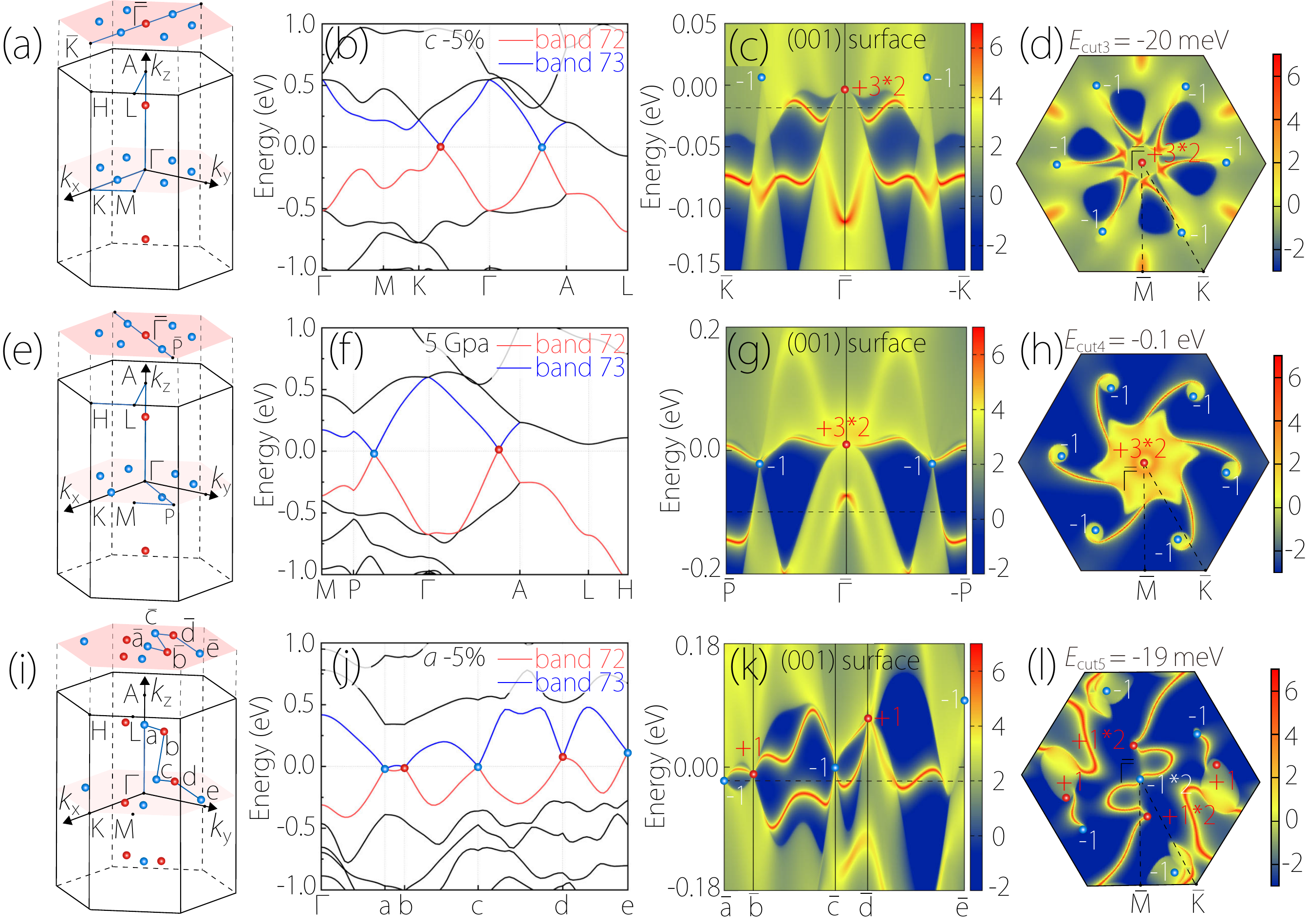}
\caption{Symmetry-controlled evolution of C-3 WPs under strain in TBR-B$_{48}$. (a–d) Evolution of the WPs, bulk band structure, surface states, and constant-energy contours under 5\% tensile strain along the $c$ axis, which preserves the crystal symmetry. (e–h) Corresponding results under 5 GPa hydrostatic compression. (i–l) Corresponding results under 5\% tensile strain along the $a$ axis, which breaks the $C_6$ symmetry. The original C-3 WPs split into three C-1 WPs, accompanied by a reconstruction of the surface Fermi arcs. The red and blue spheres represent WPs with positive and negative chirality, respectively. These results demonstrate the symmetry-controlled conversion between high-Chern-number and conventional Weyl phases.
}
\label{fig:strain}
\end{figure*}

\subsection{Chirality-dependent Weyl topology in chiral OBD-B$_{48}$}
\label{sec:obd}

To demonstrate that the symmetry-constrained $2\times\{3,-1,-1,-1\}$-IWM configuration is not restricted to a single type of hourglass-shaped Weyl connectivity, we further investigate the electronic structures of chiral OBD-B$_{48}$. Unlike TBR-B$_{48}$, where the C-3 WPs exhibit an hourglass-type connectivity, OBD-B$_{48}$ realizes a non-hourglass C-3 Weyl configuration within the same global charge-compensated framework.

We first consider the left-handed enantiomer [Fig.~\ref{fig:structure}(f)], \emph{l}-OBD-B$_{48}$, crystallizing in the chiral SG $P6_2$ (No.~171). Figure~\ref{fig:obd}(a) presents the bulk electronic structure along the high-symmetry paths. Similar to TBR-B$_{48}$, the bands around the Fermi level are clean and isolated from other electronic states, providing a suitable platform for resolving high-charge Weyl crossings. A full Brillouin-zone search of the band gap between the 72nd and 73rd bands identifies two symmetry-related WPs along the $\Gamma$-$A$ line and six additional WPs distributed along the $L$-$M$ lines.

The two WPs located along $\Gamma$-$A$ are labeled as W$_1$, while the six compensating nodes are denoted as W$_2$. Their energies, momenta, multiplicities, and topological charges are summarized in Table~\ref{tab:weylpoints}. The W$_1$ nodes belong to the representation
$\Delta_3+\Delta_4$, consistent with the C-3 Weyl symmetry channel identified in the space-group classification, whereas the W$_2$ nodes originate from the $U_1+U_2$ representation and form the compensating C-1 Weyl sector.

The projected density of states shown in Fig.~\ref{fig:obd}(b) indicates that the electronic states near the Fermi level mainly originate from strongly hybridized boron $p$ orbitals. The local band dispersion around W$_1$ [Fig.~\ref{fig:obd}(c)] exhibits the characteristic cubic behavior of a C-3 WP, whereas the corresponding dispersion around W$_2$ (Fig.~S3(b) in the SM~\cite{SM54}) confirms its conventional unit-charge Weyl character.

The topological charges are further determined from the evolution of Wannier charge centers. As shown in Fig.~\ref{fig:obd}(d), the winding of the Wannier charge centers confirms that the two W$_1$ nodes carry $C=-3$, whereas the six W$_2$ nodes carry $C=+1$. Therefore, the left-handed OBD-B$_{48}$ realizes the charge-reversed counterpart of TBR-B$_{48}$, $\{-3,-3,+1,+1,+1,+1,+1,+1\}$, while preserving the same $2\times\{3,-1,-1,-1\}$-IWM topology.

The bulk Weyl topology is directly reflected in the surface electronic structure. On the (001) surface, the two W$_1$ nodes project onto the same surface momentum, producing an accumulated projected charge of $-6$. Accordingly, the surface spectrum and constant-energy contour shown in Figs.~\ref{fig:obd}(e,f) exhibit six Fermi-arc branches connecting the projected C-3 sector to the six projected C-1 nodes, forming a sixfold helicoid Fermi-arc structure.

On the (010) surface, the two C-3 WPs project onto distinct surface momenta. As shown in Figs.~\ref{fig:obd}(g,h), each projected C-3 node connects to three Fermi arcs terminating at the projected C-1 WPs, resulting in the threefold Fermi-arc connectivity. These surface states reproduce the bulk charge distribution and further confirm the realization of the non-hourglass C-3 Weyl complex in OBD-B$_{48}$.

The role of structural chirality is further revealed by comparing the two enantiomorphic OBD-B$_{48}$ structures. The right-handed \emph{r}-OBD-B$_{48}$ structure crystallizes in the opposite chiral SG $P6_4$ (No.~172) and exhibits nearly identical electronic dispersions to \emph{l}-OBD-B$_{48}$. However, reversing the crystal handedness reverses the signs of all Weyl charges. As shown in Figs.~\ref{fig:obd}(i,l), the W$_1$ nodes in \emph{r}-OBD-B$_{48}$ carry $C=+3$, while the six W$_2$ nodes carry $C=-1$.

This charge reversal occurs without changing the number, momentum distribution, or connectivity of the WPs, demonstrating that the sign of the topological charge is directly controlled by the structural chirality.

The surface-state calculations further reveal the consequence of chirality reversal. The (001)- and (010)-surface spectra of \emph{r}-OBD-B$_{48}$ [Figs.~\ref{fig:obd}(m-p)] exhibit the same numbers of Fermi arcs as the left-handed structure, corresponding to the same projected charge magnitudes. However, the propagation directions and connectivity textures of the Fermi arcs are reversed due to the opposite Weyl charge distribution. Therefore, OBD-B$_{48}$ provides a direct material realization of the correspondence between real-space structural chirality and momentum-space topological chirality.

Together with TBR-B$_{48}$, these results demonstrate that the symmetry-defined
$2\times\{3,-1,-1,-1\}$-IWM configuration can host different microscopic realizations, including both hourglass-type and non-hourglass-type C-3 Weyl phases, while preserving the same global charge-compensated topology.

\subsection{Symmetry-controlled evolution of C-3 WPs under strain}
\label{sec:strain}

The C-3 WPs in TBR-B$_{48}$ originate from the symmetry-enforced band crossing along the $\Gamma$-$A$ direction, where the sixfold rotational symmetry constrains the interband coupling and protects the high topological charge. To further verify the symmetry origin of the C-3 topology and explore its tunability, we investigate the evolution of the WPs under different strain conditions.

We first consider two symmetry-preserving perturbations: tensile strain along the $c$ axis and hydrostatic compression. A tensile strain of 5\% along the $c$ direction modifies the lattice parameters while preserving the original $C_6$ rotational symmetry. Similarly, applying 5 GPa hydrostatic compression changes the lattice geometry without breaking the rotational symmetry. As shown in Figs.~\ref{fig:strain}(a) and \ref{fig:strain}(e), the $2\times\{3,-1,-1,-1\}$-IWM configuration remains qualitatively unchanged under both perturbations.

The corresponding band structures [Figs.~\ref{fig:strain}(b) and \ref{fig:strain}(f)] show that the Weyl crossings remain intact, with only small shifts in their energies and momenta. The surface-state calculations on the (001) surface [Figs.~\ref{fig:strain}(c) and \ref{fig:strain}(g)] further demonstrate that the sixfold Fermi-arc connectivity associated with the projected C-3 Weyl sector is preserved. The corresponding constant-energy contours [Figs.~\ref{fig:strain}(d) and \ref{fig:strain}(h)] exhibit essentially unchanged arc connectivity, confirming that the $2\times\{3,-1,-1,-1\}$-IWM topology remains stable against finite symmetry-preserving structural deformations.

These results demonstrate that the robustness of the C-3 WPs is not simply due to energetic separation of the crossings, but originates from the preservation of the crystalline symmetry that constrains their topological charge. As long as the protecting $C_6$ rotational symmetry is maintained, small continuous structural deformations cannot remove or split the C-3 WPs without closing and reopening the relevant symmetry-protected gap.

In contrast, we next consider a symmetry-breaking deformation by applying 5\% tensile strain along the $a$ axis. Unlike the previous perturbations, this deformation lowers the crystal symmetry and removes the sixfold rotational constraint protecting the C-3 WPs. Consequently, the original C-3 WP located on the $C_6$ rotation axis becomes unstable and undergoes a topological splitting transition. As shown in Fig.~\ref{fig:strain}(i), each C-3 WP splits into three C-1 WPs after the $C_6$ symmetry breaking. The total topological charge is conserved during this process, $+3\rightarrow +1+1+1$, and the compensating C-1 sector evolves accordingly to maintain the global charge-neutrality condition. The strained structure belongs to the monoclinic SG $P2_1$ (No.~4), which no longer supports the symmetry protection required for a C-3 Weyl crossing.

The reconstructed electronic structure in Fig.~\ref{fig:strain}(j) confirms the splitting of the original high-charge WPs into multiple conventional Weyl crossings. Correspondingly, the surface states shown in Fig.~\ref{fig:strain}(k) and the constant-energy contour in Fig.~\ref{fig:strain}(l) exhibit a reconstruction of the Fermi-arc connectivity. The original high-charge surface connectivity is replaced by the connectivity expected from the redistributed unit-charge WPs.

The strain-induced transition therefore provides a direct demonstration of the symmetry-controlled relationship between high-charge and conventional Weyl phases. More broadly, it illustrates that the Chern number of a WP is not only an intrinsic property of the band crossing itself, but also a consequence of the crystalline symmetry constraints imposed on the low-energy electronic structure.

\section{Discussion and Conclusion}
\label{sec:discussion}

The significance of this work extends beyond the identification of two boron-based C-3 WSMs. Previous studies of high-Chern-number Weyl fermions have mainly focused on the realization and characterization of individual high-charge WPs~\cite{BaNiIO6,SrNiIO6,La3AlCrS7,Li2CO4}. However, because WPs are constrained by the global charge-neutrality condition in a lattice, the existence of a C-3 WP is intrinsically connected to the distribution of compensating WPs throughout the BZ. In this work, we establish a symmetry-based framework for the $2\times\{3,-1,-1,-1\}$-IWM configuration, shifting the study of C-3 Weyl fermions from isolated-node identification toward a systematic understanding of charge-compensated Weyl complexes.

The predicted boron allotropes TBR-B$_{48}$ and chiral OBD-B$_{48}$ demonstrate how this symmetry framework can guide the realization of unconventional Weyl configurations in realistic materials. The weak SOC of boron preserves the crystalline-symmetry protection of the high-charge crossings, while the diverse bonding networks of boron allotropes provide structural flexibility for hosting distinct Weyl connectivities~\cite{Charge2WeylDirac,OddNodeWeylDirac,ChargeTwoTriplePoint,FourDoubleWeyl}. TBR-B$_{48}$ and OBD-B$_{48}$ realize two different manifestations of the same $2\times\{3,-1,-1,-1\}$-IWM configuration, characterized by hourglass-type and non-hourglass-type C-3 WPs, respectively. Furthermore, the opposite enantiomorphic structures of OBD-B$_{48}$ reverse the signs of all Weyl charges, revealing a direct correspondence between structural chirality and momentum-space chirality.

Beyond static realization, the strain-dependent evolution of TBR-B$_{48}$ demonstrates the tunability of high-Chern-number Weyl states through symmetry manipulation. Preserving the $C_6$ rotational symmetry maintains the C-3 WPs, whereas breaking this symmetry transforms each C-3 WP into three conventional C-1 WPs accompanied by a reconstruction of the surface Fermi arcs. This provides a direct example of how crystalline symmetry controls the conversion between high-charge and unit-charge Weyl phases.

In summary, we have established a symmetry-based framework for identifying nonmagnetic Weyl semimetals hosting the coexistence of charge-three and charge-one WPs. Through systematic space-group analysis and first-principles calculations, we identify candidate crystallographic environments and predict TBR-B$_{48}$ and chiral OBD-B$_{48}$ as material realizations of the $2\times\{3,-1,-1,-1\}$-IWM configuration. The combination of bulk topology, surface Fermi-arc structures, crystal chirality, and strain-controlled topological transitions demonstrates a general strategy for designing and manipulating rare high-Chern-number Weyl fermions in crystalline materials.

\vspace{-\baselineskip}
\begin{acknowledgments}
This work was supported by the National Natural Science Foundation of China (Grants No.~12304202), the Hebei Natural Science Foundation (Grant No.~A2023203007), the Science Research Project of Hebei Education Department (Grant No.~BJK2024085), and the Cultivation Project for Basic Research and Innovation of Yanshan University (No.~2022LGZD001).
\end{acknowledgments}

\end{document}